\documentclass[useAMS,usenatbib]{mn2e}
\usepackage{latexsym,graphicx,natbib,amssymb}

\title[Half-mass to Jacobi radii of Galactic globular clusters]
 {Evidence for two populations of Galactic globular clusters from the ratio of their half-mass to 
  Jacobi radii}
      
\author[Baumgardt et al.]
{Holger Baumgardt$^{1}$, Genevi{\`e}ve Parmentier$^{1,2}$, Mark Gieles$^3$, Enrico Vesperini$^4$\\
 $^1$ Argelander-Institut f\"ur Astronomie, Universit\"at Bonn, Auf dem H\"ugel 71, 53121 Bonn, Germany \\
 $^2$ Institute of Astrophysics \& Geophysics, University of Li\`ege, All\'ee du 6 Ao\^ut 17, B-4000 Li\`ege, Belgium \\
 $^3$ European Southern Observatory, Casilla 19001, Santiago 19, Chile \\
 $^4$ Departmentof Physics, Drexel University, Philadelphia, PA, USA  
}

\date{Accepted ????. Received ?????; in original form ?????}

\pagerange{\pageref{firstpage}--\pageref{lastpage}} \pubyear{2008}

\begin{document}   

\maketitle

\label{firstpage}
 
\begin{abstract}
We investigate the ratio between the half-mass radii $r_h$ of Galactic globular clusters and their Jacobi
radii $r_J$ given by the potential of the Milky Way and show that clusters with galactocentric distances 
$R_{GC}>8$ kpc fall into two distinct groups: one group of compact, tidally-underfilling clusters with 
$r_h/r_J<0.05$ and another 
group of tidally filling clusters which have $0.1 < r_h/r_J<0.3$. We find no correlation between the membership of
a particular cluster to one of these groups and its membership in the old or younger halo population. Based 
on the relaxation times and orbits of the clusters, we argue that compact clusters and most clusters in the
inner Milky Way were born compact with half-mass radii $r_h < 1$ pc. Some of the tidally-filling clusters 
might have formed compact as well, but the majority likely formed with large half-mass radii. Galactic globular clusters 
therefore show a similar dichotomy as was recently found for globular clusters in dwarf galaxies and for young 
star clusters in the Milky Way. It seems likely that some of the tidally-filling clusters are evolving along the main sequence line of 
clusters recently discovered by K\"upper et al. (2008) and are in the process of dissolution.
\end{abstract}


\section{Introduction}
\label{sec:intro}

It is well known that most, if not all, stars form in star clusters. Star clusters are therefore important
probes of the star formation process \citep{k05, pg09}. This is especially the case for globular clusters which, due 
to their large ages and low metallicities, are relics of star formation processes in the early universe.

Observations show that in nearby galaxies star clusters form compact, with half-mass radii of $r_h<1$ pc, 
at the centres of giant molecular cloud cores \citep{ll03}. They then undergo a significant expansion as a
result of gas expulsion driven by stellar winds and UV radiation from bright stars and supernovae explosions
\citep{h80, bg06, bk07} and later also by mass loss from stellar evolution \citep{cw90, fh95, vz03}. As a result, the 
radii of star clusters show a steady increase with cluster age within the first 20 Myrs \citep{betal08,p09}.

Globular clusters are dynamically evolved systems, so in addition to 
the early evolution due to gas expulsion and stellar evolution, cluster radii are also subject to dynamical 
cluster evolution due to two-body relaxation. As a result of mass segregation and core collapse, the 
core radius shrinks while the half mass 
radius stays roughly constant before core collapse. After core collapse, stellar binaries provide a central 
heat source and the cluster expands self-similarly \citep{g84,metal90,getal91,gh94,bhh02,hetal06}. This process continues until the cluster 
runs into the external tidal field, at which point cluster expansion is balanced by the loss of outer stars over the
tidal boundary. As a result, the ratio of half-mass radius to Jacobi radius\footnote{Throughout the paper we will
denote the distance from the centre of a star cluster to the first Lagrangian point as the Jacobi radius $r_J$, while
the term tidal radius $r_t$ refers to the limiting radius of King (1962) or King (1966) models.}
evolves along a 
common sequence which, at least for single-mass clusters, only depends on current cluster mass \citep{kkb08}. 
The ratio of the half-mass to Jacobi radius for 
individual clusters therefore contains important information on the dynamical state of a star cluster.

In the current paper we examine the distribution of half-mass to Jacobi radii of Galactic globular clusters
in order to better understand the formation and evolution of Galactic globular clusters.
The paper is organised as follows: In Sec.~\ref{sec:gcd} we present the data and show that outer globular
clusters fall into two distinct groups. In Sec. \ref{sec:disc} we discuss the relation of 
these groups to different subsamples of Galactic globular clusters like old and younger
halo clusters or globular clusters believed to be accreted from dwarf galaxies and use additional information 
like half-mass relaxation times and cluster orbits to determine the degree of dynamical cluster evolution.
In Sec. \ref{sec:concl} we finally draw our conclusions.

\section{Globular cluster data}
\label{sec:gcd}

In order to investigate the degree of tidal filling, one has to obtain an estimate of the Jacobi
radius of a cluster. Most investigations so far used the tidal radius $r_t$ obtained by 
fitting the observed surface density profile with an empirical profile like \citet{k62} or
a theoretical one like \citet{k66} as an estimate of the Jacobi radius.
Surface density data for most globular clusters is however either not available near the Jacobi
radius or becomes unreliable in the outer parts due to the low number of cluster stars and the uncertain density of
background stars. In addition, due to the high number of stars, surface densities 
can be much better determined in the inner cluster parts, so that the published tidal radii $r_t$ 
of globular clusters are determined more by the density
profile inside a few half-mass radii and might not reflect the true tidal radius of a cluster.
\citet{betal09} for example found that in case of NGC~2419, the best-fitting King model has a nominal tidal radius
of 150 pc, while they estimated that the Jacobi radius $r_J$ of the cluster is around 800 pc.

\begin{figure}
\begin{center}
\includegraphics[width=8.5cm]{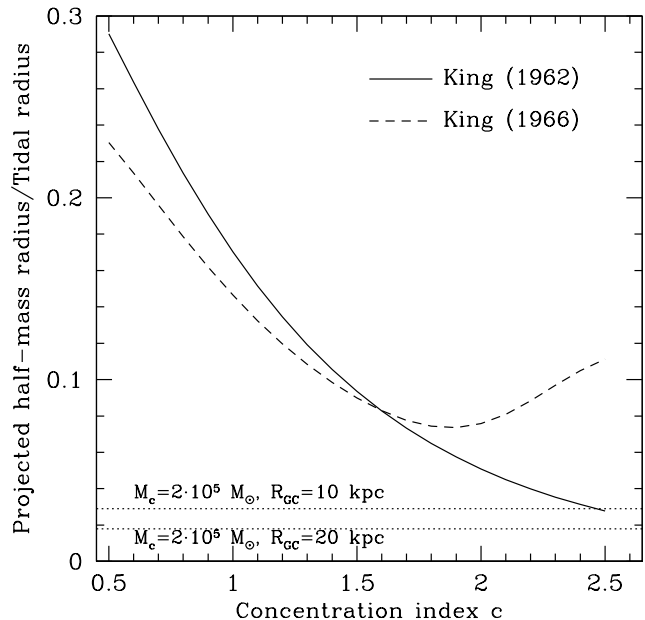}
\end{center}
\caption{Ratio of projected half-mass radius $r_{hp}$ to tidal radius $r_t$ as a function of the
concentration index $c=\log_{10} r_t/r_c$ for King (1962) and King (1966) models. The dotted lines 
mark expected ratios of $r_{hp}/r_J$ for a typical globular cluster with $M_c=2 \cdot 10^5$ M$_\odot$ and
$r_{hp}=3$ pc at $R_{GC}=10$ kpc and $R_{GC}=20$ kpc galactocentric distance. The $r_{hp}/r_J$ ratio of 
such a cluster cannot be reproduced by any King model, showing that fitting King models to observed 
density profiles in order to derive the Jacobi radius can lead to a significant bias towards a too small 
value.}
\label{fig:king}
\end{figure}

An additional problem of fitting King models to determine Jacobi radii is depicted in Fig.\ \ref{fig:king}.
This figure shows the ratio between the projected half-mass radius and the tidal radius $r_{hp}/r_t$ for \citet{k62} and
\citet{k66} models of various concentrations $c=\log_{10}{r_t/r_c}$. It can be seen that with both families of models
only a limited range of $r_{hp}/r_t$ values can be reached. \citet{k62} models
are restricted to values between $0.028 \le r_{hp}/r_t \le 0.29$ while \citet{k66} models are restricted to the
even smaller range $0.074 \le r_{hp}/r_t \le 0.23$. A typical globular cluster with a mass of $M_c=2 \cdot 10^5$~M$_\odot$ 
and a projected half-mass radius of $r_{hp}=3$ pc has $r_{hp}/r_J=0.029$ at a galactocentric distance of $R_{GC}=10$ kpc 
and $r_{hp}/r_J=0.018$ at $R_{GC}=20$ kpc. Hence King models cannot accurately describe the density profile of such a cluster
and one would have to use models allowing for more extended envelopes like those described by \citet{w75}
to fit the outer surface density profile. This was also noted by \citet{mm05}, who found that Wilson models provide 
equally good or significantly better fits than King (1966) models for about 90\% of their sample of young massive 
clusters and old globular clusters. 

In the present paper we use the ratio of the half-mass radius to the Jacobi radius $r_J$ as a measure 
for the degree of tidal 
filling that a globular cluster experiences in the tidal field of the Galaxy. The Jacobi radius can be determined 
if the cluster mass, galactocentric distance and the underlying Galactic potential 
is known. For typical globular clusters, masses and galactocentric distances should have typical errors of less
than 20\%, hence Jacobi radii can for many clusters be determined with higher accuracy
than their tidal radii $r_t$. The main disadvantage is that the Jacobi radius varies along the orbit of a 
cluster, so for highly eccentric orbits, the current radius might not be a good measure for the average Jacobi 
radius which a cluster experiences and which determines its mass loss rate. We will discuss the influence 
of eccentric orbits on our results further below. 

In order to calculate the ratio $r_h/r_J$ of Galactic globular clusters, we calculated 
3D half-mass radii $r_h$ from the projected half-light radii $r_{hp}$ under the assumption 
that mass follows light and by assuming $r_h = 1.33 r_{hp}$. This relation is correct to
within 5\% for most \citet{k62} or \citet{k66} density profiles. We note that half-light radii
can be different from half-mass radii for highly evolved clusters that have undergone core-collapse
\citep{bm03, baetal09}. However, correcting for this effect would require detailed observational data
for each cluster which is not available at the moment. Jacobi radii were calculated according to
\citet{k62} (see also Innanen, Harris \& Webbink 1983 who added a factor of 2/3 to correct for the 
elongation in the direction along the line connecting the Lagrangian points):
\begin{equation}
     r_J = \left( \frac{G \; M_c}{2 \; V_G^2} \right)^{1/3} \;\; R_{GC}^{2/3}  \;\; .
\label{rtide}
\end{equation}
Here $M_c$ is the mass of the cluster, $V_G$ the circular velocity of the galaxy
and $R_{GC}$ the distance of the cluster from the galactic centre. We assumed a spherically symmetric
density distribution for the Milky Way with a constant circular velocity of $V_G=220$ km/sec.
Cluster masses were calculated from the absolute luminosities of the clusters and an assumed V-band 
mass-to-light ratio of $M/L_V=2.0$.
The cluster data (projected half-light radii, total luminosities, galactocentric distances)
was taken from the 2003 version of the Globular Cluster database of \citet{h96}, supplemented by
additional data for a few clusters from \citet{bb08}. We exclude Omega~Cen, M54 and NGC~2419
from our analysis since these clusters might be stripped nuclei of dwarf galaxies rather than 
genuine globular clusters.

\begin{figure}
\begin{center}
\includegraphics[width=8.5cm]{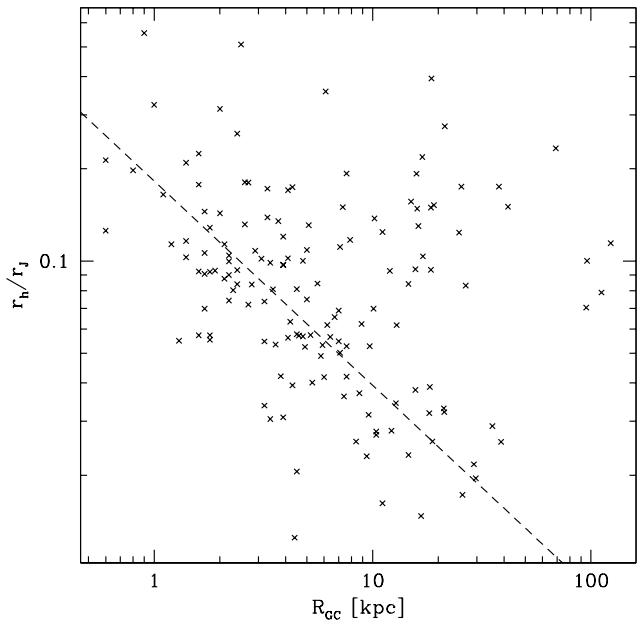}
\end{center}
\caption{Ratio of half-mass radius $r_h$ to Jacobi radius $r_J$ as a function of galactocentric
distance $R_{GC}$ for Galactic globular clusters. It can be seen that clusters outside $R_{GC}
\approx 8$ kpc fall into two distinct groups, clusters with $r_h/r_J<0.05$ and clusters with
$0.07 < r_h/r_J<0.3$. The dashed line depicts the position that a $M_c=10^5$ M$_\odot$ with
$r_h=3$ pc would have at different galactocentric distances. Compact clusters outside 8 kpc 
and most clusters inside this radius fall onto this line.}   
\label{fig:rhrj}
\end{figure}

Fig.\ \ref{fig:rhrj} depicts the ratio of 3D half-mass radius $r_h$ to Jacobi radius $r_J$ as a function of
galactocentric distance. It can be seen that the distribution of Galactic globular clusters is not uniform in this
plane. First, clusters with large values of $r_h/r_J>0.5$ are basically absent. Since such clusters would be subject
to strong tidal forces and have consequently small dissolution times, if they ever existed they should be quickly 
destroyed and not be present any more after 10 Gyr of evolution. The absence of
such clusters in our distribution therefore serves as a sanity check of our method. 
%%First, a number of globular clusters inside $R_{GC}=2$ kpc have
%%relatively large values of $r_h/r_J$ between 0.3 up to nearly 0.6. If correct, these values would imply 
%%that these clusters are subject to strong tidal forces and therefore have very short lifetimes, 
%%making it unlikely to find such clusters after a Hubble time of evolution. An alternative explanation
%%is that these clusters are on eccentric orbits and are located at much larger galactocentric 
%%distances for most of their lifetime. Also observational errors in distance could have lead
%%to wrong values of $R_{GC}$ for these clusters. Distance errors are especially important if the line-of-sight
%%passes close to the galactic centre since in such a case a small relative error in distance to the sun can lead 
%%to a much larger relative error in galactocentric distance. 

Second, most clusters inside $\sim8$~kpc exhibit a relatively broad distribution of $r_h/r_J$ values between $0.02 < r_h/r_J < 0.2$
without any noticeable separation. Outside about 8 kpc, the Galactic globular clusters can be split into two groups, one group of
clusters with $r_h/r_J<0.05$ and a second group of clusters with $0.08 < r_h/r_J<0.3$. A KS test
gives only an 14\% chance that clusters beyond 8 kpc follow a log-normal distribution in $\log r_h/r_J$.
In addition, the average mass of extended clusters in the outer Milky Way is significantly lower
than the mass of the more compact clusters (see Fig.\ 4), the mean mass of compact clusters is
$\log{M_c} = 5.44 \pm 0.11$ while the extended group has a mean mass of only $\log{M_c} = 4.38 \pm 0.06$.
Both results indicate that two distinct groups of globular clusters exist in the Milky Way. The dividing line between
both groups seems to be around $r_h/r_J=0.07$ and Table~1 lists the basic parameters of clusters having
$r_h/r_J$ ratios smaller or larger than this value. Different orbits seem unlikely
to be an explanation for this dichotomy since clusters of both groups are located within the same interval of 
galactocentric distances and, at least for those clusters with orbital information, the average ratios of $R_{GC}/R_{Peri}$
are similar.

The dashed line in Fig.~\ref{fig:rhrj} depicts the position which a $M_c=10^5$ M$_\odot$ with $r_h=3$ pc would have in the
$r_h/r_J$ vs. $R_{GC}$ plot at different galactocentric distances. It can be seen that clusters in the lower 
group outside 8 kpc and most clusters inside this radius fall onto this line. Most clusters in the lower
group are therefore relatively massive and compact, something which can also be seen in Table~1. Clusters in the upper
group strongly feel the tidal field of the Galaxy. It can be seen that this group is more diverse since
it is made up of massive and extended as well as low-mass compact clusters. In the following, we will 
discuss possible reasons for the origin of both groups.

\begin{table*}
\caption[]{Basic data for Galactic globular clusters with $R_{GC}>8$ kpc belonging to both groups}
\begin{tabular}{lrccrrr|lrccrrr}
\hline\hline
\multicolumn{1}{c}{Name} & \multicolumn{1}{c}{$r_h$} &\multicolumn{1}{c}{$r_h/r_J$} &  \multicolumn{1}{c}{$\log M_C$} & \multicolumn{1}{c}{$\log T_{RH}$} &
  \multicolumn{1}{c}{$R_{GC}$} & \multicolumn{1}{c}{$R_{Peri}$} & 
\multicolumn{1}{c}{Name} & \multicolumn{1}{c}{$r_h$} & \multicolumn{1}{c}{$r_h/r_J$} & \multicolumn{1}{c}{$\log M_C$} & \multicolumn{1}{c}{$\log T_{RH}$} &
  \multicolumn{1}{c}{$R_{GC}$}  & \multicolumn{1}{c}{$R_{Peri}$} \\ 
 &  \multicolumn{1}{c}{[pc]} & & \multicolumn{1}{c}{[$M_\odot$]} &  \multicolumn{1}{c}{[yr]} &  
\multicolumn{1}{c}{[kpc]} & \multicolumn{1}{c}{[kpc]} & & 
 \multicolumn{1}{c}{[pc]} &  & \multicolumn{1}{c}{[$M_\odot$]} &  \multicolumn{1}{c}{[yr]} & \multicolumn{1}{c}{[kpc]} 
& \multicolumn{1}{c}{[kpc]} \\
\hline 
\multicolumn{6}{c}{Compact group} &\multicolumn{6}{c}{Tidally filling group} \\
NGC  362 &     2.67 &    0.023  &    5.60 &     9.09 &      9.4 & 0.8  & NGC  288 &     7.58 &    0.093 &     4.93 &     9.50 &     12.0 & 5.3  \\ 
NGC 1261 &     4.77 &    0.032  &    5.36 &     9.37 &     18.2 &      & Pal    1 &     2.87 &    0.104 &     3.22 &     8.23 &     17.0 &      \\ 
Pal    2 &     7.17 &    0.029  &    5.44 &     9.67 &     35.4 &      &  AM    1 &    23.64 &    0.114 &     4.12 &     9.92 &    123.2 &      \\ 
NGC 1851 &     2.44 &    0.015  &    5.57 &     9.02 &     16.7 & 5.7  & Eridanus &    13.99 &    0.070 &     4.29 &     9.65 &     95.2 &      \\ 
NGC 1904 &     4.00 &    0.026  &    5.38 &     9.27 &     18.8 & 4.2  & Pyxis &    20.79 &    0.150 &     4.53 &    10.00 &     41.7 &      \\ 
NGC 2298 &     3.24 &    0.038  &    4.75 &     8.87 &     15.7 & 1.9  & Pal    3 &    23.73 &    0.100 &     4.51 &    10.08 &     95.9 & 82.5 \\ 
NGC 2808 &     2.83 &    0.016  &    5.99 &     9.29 &     11.1 & 2.6  & Pal    4 &    22.87 &    0.079 &     4.64 &    10.10 &    111.8 &      \\  
NGC 3201 &     5.20 &    0.062  &    5.22 &     9.37 &      8.9 & 9.0  & Rup  106 &     9.04 &    0.094 &     4.77 &     9.55 &     18.5 &      \\ 
NGC 4147 &     3.22 &    0.032  &    4.70 &     8.85 &     21.3 & 4.1  & NGC 5053 &    22.26 &    0.218 &     4.92 &    10.20 &     16.9 &      \\ 
NGC 4590 &     6.13 &    0.070  &    5.17 &     9.46 &     10.1 & 8.6  &  AM    4 &     4.87 &    0.175 &     2.87 &     8.46 &     25.5 &      \\ 
NGC 5024 &     7.66 &    0.039  &    5.71 &     9.83 &     18.3 & 15.5 & NGC 5466 &    13.87 &    0.130 &     5.02 &     9.93 &     16.2 & 5.4  \\ 
NGC 5272 &     4.52 &    0.028  &    5.81 &     9.52 &     12.2 &  5.5 &  IC 4499 &    11.00 &    0.094 &     5.17 &     9.84 &     15.7 &      \\ 
NGC 5286 &     2.94 &    0.026  &    5.68 &     9.19 &      8.4 &      & Pal    5 &    26.63 &    0.395 &     4.30 &    10.07 &     18.6 & 6.1  \\ 
NGC 5634 &     5.28 &    0.033  &    5.31 &     9.42 &     21.2 &      & Pal   14 &    32.96 &    0.233 &     4.13 &    10.14 &     69.0 &      \\ 
NGC 5694 &     4.44 &    0.022  &    5.36 &     9.32 &     29.1 &      & NGC 6101 &    10.15 &    0.124 &     5.00 &     9.72 &     11.1 &      \\ 
NGC 5824 &     4.47 &    0.017  &    5.77 &     9.50 &     25.8 &      & Pal   15 &    20.93 &    0.175 &     4.43 &     9.96 &     37.9 &      \\ 
NGC 6205 &     4.45 &    0.037  &    5.71 &     9.47 &      8.7 & 5.0  & NGC 6426 &     7.71 &    0.084 &     4.91 &     9.50 &     14.6 &      \\ 
NGC 6229 &     4.36 &    0.020  &    5.45 &     9.35 &     29.7 &      & Ter    7 &     8.73 &    0.148 &     4.25 &     9.33 &     16.0 &      \\ 
NGC 6341 &     3.47 &    0.031  &    5.51 &     9.23 &      9.6 & 1.4  & Arp    2 &    21.19 &    0.276 &     4.35 &     9.94 &     21.4 &      \\ 
NGC 6779 &     4.54 &    0.053  &    5.19 &     9.27 &      9.7 & 0.9  & Ter    8 &    10.08 &    0.152 &     4.25 &     9.42 &     19.1 &      \\ 
NGC 6864 &     3.77 &    0.023  &    5.65 &     9.34 &     14.6 &      & Pal   12 &     9.48 &    0.193 &     4.03 &     9.29 &     15.9 &      \\ 
NGC 6934 &     3.65 &    0.034  &    5.22 &     9.14 &     12.8 & 6.0  & Pal   13 &     4.60 &    0.083 &     3.73 &     8.71 &     26.7 &      \\ 
NGC 6981 &     5.80 &    0.062  &    5.05 &     9.37 &     12.9 &      & NGC 7492 &    12.21 &    0.124 &     4.54 &     9.66 &     24.9 &      \\ 
NGC 7006 &     6.12 &    0.026  &    5.31 &     9.51 &     38.8 & 18.2 &  IC 1257 &    13.57 &    0.149 &     4.69 &     9.79 &     18.5 &      \\ 
NGC 7078 &     4.23 &    0.027  &    5.90 &     9.52 &     10.4 &  5.4 &  BH  176 &     4.84 &    0.138 &     3.97 &     8.84 &     10.2 &      \\ 
NGC 7089 &     4.15 &    0.028  &    5.84 &     9.48 &     10.4 &  6.4 & ESO  280 &     8.42 &    0.156 &     4.19 &     9.28 &     15.0 &      \\ 
\hline \hline
\end{tabular}
\label{tabgcdata}
\end{table*}

\section{Discussion}
\label{sec:disc}

\subsection{Correlation with old and younger halo membership}
\label{sec:dsc-halo}

The Galactic Globular cluster system consists of different subsystems.
While bulge/disc GCs differ from halo clusters with respect
to their metallicity ([Fe/H] $\geq -0.8$ and [Fe/H] $< -0.8$, respectively), 
the halo subsystem itself is made of clusters with more than one origin.
It is traditionally splitted up into two groups, referred to as the Old Halo and 
the Younger Halo (Zinn 1993, van den Bergh 1993, Mackey \& Gilmore 2004), based 
on differences in horizontal branch (HB) morphology, age, kinematics, 
spatial distribution (see Parmentier et al.~2000, their Section 2, for a review).  
Because of their predominant location beyond the Solar Circle,
YH clusters are assumed to have been accreted -- along with the dwarf galaxies 
which used to host them -- after the main body of the Galaxy was built up.  
Depending on how late they were accreted into the Galactic halo, their 
evolutionary history may be different from what {\it in-situ} OH clusters have
experienced in the Milky Way tidal field.  The current accretion of the Sagittarius 
dwarf galaxy and of its small globular cluster system is the smoking gun of this process (Ibata, Gilmore \& Irwin 1994). 

\begin{figure}
\begin{center}
\includegraphics[width=8.5cm]{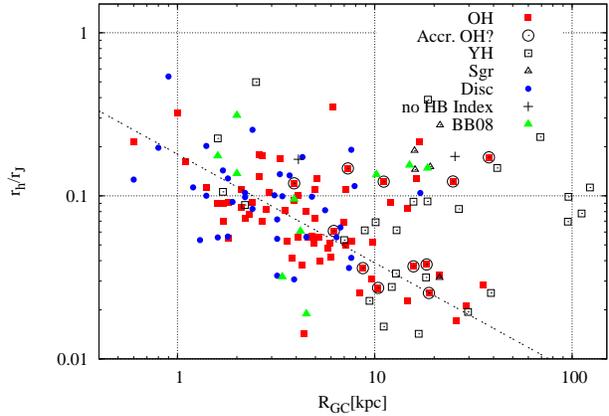}
\end{center}
\caption{Ratio of $r_h/r_J$ vs. galactocentric distance for Galactic globular clusters. Different symbols
indicate to which subsystem a cluster belongs, old halo clusters are marked by filled squares, younger halo clusters by open squares, 
disc globular clusters by filled circles and clusters accreted from the Sagittarius dwarf galaxy by open triangles. There is no 
correlation between the classification of a globular cluster and its $r_h/r_J$ value.}
\label{fig:ohyh}
\end{figure}

In Fig.~\ref{fig:ohyh}, data points are symbol-coded to highlight these different
cluster origins.  Filled circles depict disc GCs ([Fe/H] $\geq -0.8$), open triangles
show GCs associated to the merging dwarf galaxy Sagittarius (Ter7, 
Arp2, Ter8, Pal12, NGC4147; see Da Costa \& Armandroff 1995, 
Martinez-Delgado et al.~2002 and Bellazzini et al.~2003 for cluster membership).
Plus-signs stand for GCs with no HB morphology index.
A list of OH clusters (filled squares) is provided
in Parmentier \& Grebel (2005, their Table 1).  Other clusters are sorted in the 
YH group (open squares).  Mackey \& Gilmore (2004) emphasize that accreted clusters
could also contribute a small fraction of the OH component.  Based on either large
core radius reminiscent of those observed for GCs in satellite galaxies (their fig.~16)
or spatial motions more typical of YH objects (see also Dinescu et al.~1999), 
they identify 11 OH GCs which might have been accreted (NGC~6809, 6101, 7492, 5897 
and Pal~15 in the first category and NGC~1904, 2298, 5024, 5904, 6205, 7089 in the 
second category).  These ill-defined status clusters are shown as filled squares with open circles 
in Fig.~\ref{fig:ohyh}. Clusters from Bonatto \& Bica (2008) finally are shown as filled triangles.\\

It can be seen that there is no correlation among the classification of a cluster to either younger or
old halo group and its $r_h/r_J$ value. The compact cluster group contains 11 younger halo clusters 
and 10 old halo clusters unsuspected of having been accreted. They represent each $\sim$ 40\,\% of the 
total number of clusters (26) in the compact cluster group. The tidally filling cluster group
contains 10 younger halo clusters and 5 old halo clusters unsuspected of having been
accreted. The higher fraction of younger halo clusters, however, is mostly driven
by the 4 clusters with $R_{GC} \sim 100$\,kpc. Considering the same radial extent
as for the compact group, that is, $R_{GC}=8-50$\,kpc, younger and old halo clusters contribute
similarly to the tidally filling group, with 6 and 5 clusters, respectively, out of 22 clusters. 
Corresponding number fractions ($\sim$ 25\,\%) agree with previous ones within the
statistical uncertainties.
Moreover, both compact and tidally filling groups are characterized by the same number fraction
of clusters accreted or suspected of having been accreted (i.e. younger halo clusters or Sagittarius clusters or old halo
clusters suspected of having been accreted - see above), namely, $\sim$60\%.
We therefore conclude that the rh/rJ dichotomy is not due to a different origin of the two cluster populations 

\subsection{Origin of the compact cluster group}
\label{sec:dsc-comp}

Tab.\ \ref{tabgcdata} lists the basic parameters of clusters belonging to either group, including the
3D half-mass radius, current mass and galactocentric distance. 
It also shows the current relaxation time, calculated according to \citet{s87}:
\begin{equation}
T_{RH} = 0.138 \frac{\sqrt{M_c} r_h^{3/2}}{\sqrt{G} <\!\!m\!\!> \ln 0.11 M_c/<\!\!m\!\!>}
\end{equation} 
where $<\!\!m\!\!>=0.4 \mbox{M}_\odot$ is the average mass of stars and $G$ the gravitational constant. It can 
be seen that the compact
clusters mostly have very large relaxation times. The average relaxation time for a cluster in this
group is about 2.8 Gyr, and nearly all clusters have relaxation times larger than 1 Gyr. According to \citet{gfr04},
it takes about 7 to 10 initial half-mass relaxation times until star clusters with a narrow mass spectrum
where the massive stars are about twice as massive as the average cluster star, which is typical
for globular clusters, have gone into core collapse. The compact clusters should therefore still be 
mostly in their pre-core collapse phase and should not have started post-core collapse expansion. 

Tab.\ \ref{tabgcdata} also lists the perigalactic distances of the globular clusters as determined by 
\citet{detal99}, \citet{amp06} and \citet{detal07}. Although there are clusters which have perigalactic
distances of less than 2 kpc, for the majority of the compact group clusters, the perigalactic distances 
are within a factor of 3 of the current galactocentric distance. The estimates of $r_J$ would therefore
decrease by no more than a factor of 2 if we used the perigalactic distance to calculate $r_J$.
Most compact clusters therefore have $r_h/r_J<0.1$ also at perigalacticon and are at most moderately 
influenced by the Galactic tidal field. Hence, their small half-mass radii are likely not due to tidal 
stripping at perigalacticon but must have been the result of the formation process. 

We conclude that clusters in the compact group also formed very compact. $N$-body simulations show that the expansion 
factor due to gas expulsion is typically a factor 2 to 3 for moderate star formation efficiencies of 30\% to
40\% \citep{bk07} and 
stellar evolution will increase this value by another factor 2 if mass is lost adiabatically. Hence the initial half-mass 
radius of clusters in the compact group must have been around 1 pc or less. Since most star clusters inside 8 kpc have 
half-mass radii very similar to compact group star clusters, it seems likely that most globular clusters in the Milky 
Way formed compact and with half-mass radii of 1~pc or less, which is comparable to the half-mass radius of embedded 
star clusters in the Milky Way \citep{ll03}.
 
\subsection{Origin of the tidally filling cluster group}
\label{sec:dsc-loos} 

\begin{figure}
\begin{center}
\includegraphics[width=8.5cm]{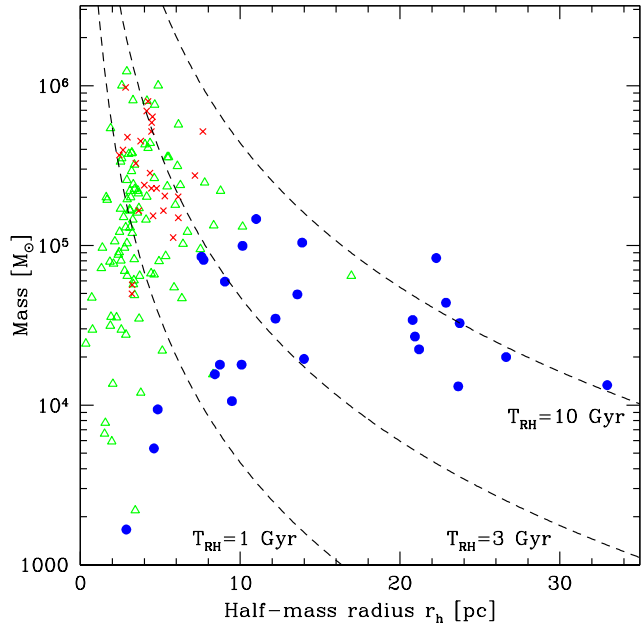}
\end{center}
\caption{Half-mass radius $r_h$ vs. cluster mass for globular clusters with $R_{GC}<8$ kpc (green triangles) and for clusters
$R_{GC}>8$ kpc that are weakly influenced by the
Galactic tidal field ($r_h/r_J<0.07$, red crosses) and for strongly tidally influenced clusters
($0.07 < r_h/r_J<0.3$, blue dots). Clusters weakly influenced by the tidal field are all massive and compact
while strongly tidally influenced clusters have significantly smaller masses. Inner clusters also have smaller 
masses on average, which might be a result of their stronger dissolution. Dashed lines show where clusters with
a given half-mass relaxation time are located in this plot.}
\label{fig:mvsrh}
\end{figure}

Fig.\ \ref{fig:mvsrh} depicts the position of inner clusters (green triangles) and of clusters with $r_h/r_J<0.07$ 
(red crosses) and clusters
with $r_h/r_J>0.07$ (blue dots) in a half-mass radius vs. mass diagram. It can be seen that clusters
with $r_h/r_J<0.07$ are mostly massive clusters with half-mass radii of a few pc, while clusters with
$r_h/r_J>0.07$ have larger radii and also smaller masses. Due to the smaller masses, clusters in the tidally filling 
group should on average be closer to dissolution. This is confirmed by observational data for a few clusters 
like Pal~5, which has very pronounced tidal tails and might be on its final orbit before dissolution 
\citep{oetal01,detal04}. Clusters in the inner Milky Way also have smaller masses than compact outer
clusters which might be due to stronger cluster dissolution in the inner Milky Way as a result
of the stronger tidal field \citep{vh97, bm03}.\\ 

One way to explain the large radii of the tidally filling clusters would be that they also formed extended. Indeed,
\citet{e08} has recently discussed different modes of star formation and attributed the difference between star 
formation in bound clusters and loose groupings to a difference in cloud pressure and different background tidal
forces. This could explain why clusters with low densities are only found far away from the centers of major galaxies 
or in dwarf galaxies. The fact that Milky Way globular clusters are clearly separated in $r_h/r_J$ is however 
more difficult to understand if cluster radii are set at formation time. An alternative viewpoint would be that the 
tidally filling clusters expanded from smaller radii, possible e.g. through post-collapse expansion driven by a population 
of stellar binaries in the cluster core.
\citet{g84} and \citet{bhh02} (their Eq.\ 4) estimated that during post-core collapse expansion, the half-mass 
radius of an isolated cluster satisfies
\begin{equation}
r_h(t) = r_{h0} \left(t/t_{cc}\right)^{(2+\nu)/3}
\end{equation}
where $r_{h0}$ is the initial half-mass radius, $t_{cc}$ the time of core collapse and $\nu \approx 0.1$ a constant
related to the cluster mass loss. \citet{gb08} found that the above relation also holds for clusters in a tidal field
as long as $r_h/r_J<0.05$. For clusters with a narrow mass spectrum, core collapse happens after 7 to 10 initial 
half-mass relaxation times \citep{gfr04}, in which case the above relation would predict that expanding clusters should have 
relaxation times which are roughly 1/10th of their current age, i.e. of order $T_{RH} \approx 10^9$ yrs. The
majority of clusters in the tidally filling group however have relaxation times $T_{RH} > 3 \cdot 10^9$ yrs, which is
too large to be explained by binary driven expansion from small radii. Also the fact that tidally filling clusters have
on average larger relaxation times than compact clusters argues against post-collapse expansion from smaller radii.
  
\citet{metal04} and \citet{metal07} have
shown that stellar mass black holes, if present in sufficient numbers, can cause strong cluster expansion.
For clusters retaining all the black holes formed in them, \citet{metal08} found that the core radius can reach values
up to 8 pc after 10 Gyr of evolution and is almost as large as the half-mass radius. This value is large enough to explain
the half-mass radii of a significant fraction of clusters in the tidally filling group (see Fig.~\ref{fig:mvsrh}).
Interestingly, in such a case clusters of the tidally filling group would have been the most compact clusters initially 
such as to be able to retain their BHs. However, some clusters in the tidally-filling group have half-mass
radii too large to be explained by BH-driven expansion and there is no significant difference in metallicity 
between compact and tidally filling clusters, as might be expected if BH kick velocities depend on metallicity, which
both argue against BH driven expansion. Central intermediate-mass black holes can also act as an efficient heat 
source, but judging from the results of \citet{bme04}, the half-mass radii of most 
clusters in the tidally filling group are too large to be explained by intermediate-mass black hole driven expansion.
Strong expansion is also possible by stellar evolution if star clusters are initially mass segregated 
since the fractional loss of potential energy can in such a case be much larger than the mass fraction lost by stellar 
evolution \citep{vetal09}.

The question of whether clusters in the tidally filling group were born compact and later expanded or already
formed with the large half-mass radii we see today therefore remains open. If they formed with large half-mass radii,
their initial relaxation times were also quite large and the clusters should be dynamically less evolved.
In this case they would not be mass segregated, so measuring stellar mass functions at different radii might be one 
way to test the formation scenario. In this context it is interesting to 
note that Jordi et al. (2009) recently found that the stellar mass function of Pal~14, which is one of the clusters 
with the longest relaxation time in our sample, differs from a Kroupa IMF inside the clusters half-mass radius.
Unfortunately no information on the stellar mass function in the outer cluster parts is available at the moment
to test whether this is due to dynamical cluster evolution.

We note that the tidally filling clusters, due to the large half-mass radii and galactocentric distances of many of them,
are partly responsible for driving the correlation between $r_h$ and $R_{GC}$ found by \citet{mv05}. A linear least-square
fit gives a relation between half-mass radius and galactocentric distance $\log{r_h}=0.50 \cdot \log{R_{GC}}+0.27$ 
for all Galactic GCs. This relation flattens to
$\log{r_h}=0.25 \cdot \log{R_{GC}}+0.38$ if we exclude the tidally filling clusters, indicating that globular clusters in the
compact group formed with similar parameters nearly everywhere in the Galaxy.

\section{Conclusions}
\label{sec:concl}

We have studied the distribution of the ratio of half-mass radius $r_h$ to Jacobi radius $r_J$ for Galactic globular 
clusters and have shown that clusters with distances larger than $\sim 8$ kpc fall into two distinct groups: One group of
compact clusters with $r_h/r_J<0.05$ and a group of more extended clusters with $0.08 < r_h/r_J<0.30$. Compact
clusters are mainly massive clusters with half-mass radii of a few pc. The half-mass radius and density of the compact 
clusters in the outer halo seems not to be adjusted to the Jacobi radius, so they were probably also born compact with 
half-mass radii $r_h<1$ pc, comparable to the half-mass radii of embedded clusters and young open clusters
in the Milky Way. Tidal radii derived from fitting King profiles to the surface density profiles of these
clusters can be significantly smaller than their Jacoby radii since the $r_h/r_J$ ratios of these clusters are smaller 
than what can be reached with any King profile.

Some of the tidally filling clusters might also have formed compact and could have 
expanded later due to dynamical heating by binary stars, stellar-mass black holes or 
intermediate-mass black holes, although it is unclear if this holds for all clusters in this group since 
about half of the tidally filling clusters have relaxation times of the order of a Hubble time or larger.

\citet{dc09} have recently found a bi-modality of the globular cluster size 
distribution in dwarf galaxies. The average radii of clusters in both of their groups agree 
quite well with the radii 
of Milky Way globular clusters in our compact and tidally filling group, showing that globular clusters formed
under similar conditions in different galaxies.
Furthermore, \citet{p09} has recently shown that open clusters in the Milky Way evolve along two sequences in 
the age vs.\
radius plane, one group of clusters starting compact with half-mass radii $r_h<1$ pc and reaching sizes of
a few pc after 20 Myr of evolution and a second group of clusters starting with half-mass radii larger than a few pc and
reaching $\sim 20$ pc after 20 Myr. The latter value agrees quite well with the sizes of most
clusters in the tidally filling group. Galactic globular clusters therefore show the same dichotomy
seen for globular clusters in dwarf galaxies and for young star clusters in the Milky Way.
 Extended star clusters appear therefore as an ubiquitous
feature of star cluster systems hosted by a variety of galaxies.
It would be interesting to see how the extended globular
clusters of the Milky Way relate  to other star clusters with large
$r_h$ values, like the Faint Fuzzy star clusters in lenticular
galaxies \citep{lb00}, the diffuse star clusters
found by \citet{petal06} in early-type galaxies of the Virgo cluster
and those hosted by dwarf galaxies \citep{dc09}.

\citet{kkb08} found that initially compact star clusters in a tidal
field expand after core collapse until they reach a mass-dependent $r_h/r_J$ value and then evolve
along a common sequence towards dissolution. They dubbed the latter phase the main sequence evolution of
star clusters. During the main sequence phase, the $r_h/r_J$ values
increase slowly with decreasing cluster mass. Extrapolating from the results of \citet{kkb08} to $M_c=10^5$ M$_\odot$, we
expect that globular clusters should have approximately $r_h/r_J \approx 0.1$ when on the main sequence,
which fits observed $r_h/r_J$ of clusters in the tidally filling group rather well. It is therefore likely
that part of the clusters in the tidally filling group, especially those with small relaxation times, have reached
the main-sequence stage of their evolution and are evolving towards dissolution. As for those with long relaxation
times, however, whether their large half-mass radius is an imprint of their formation process or a result of 
cluster expansion remains an open question.  

We finally note that it is possible that the extended clusters were initially much more numerous, since 
due to their large sizes, they are effectively destroyed by the Galactic tidal field, especially in the inner 
part of the Milky Way.

\section*{Acknowledgements}

We thank the referee for comments which improved the presentation of the paper. HB acknowledges support from the 
German Science Foundation through a Heisenberg fellowship. GP acknowledges support 
from the Belgian Science Policy Office in the form of a Return Grant and from the Alexander von Humboldt Foundation 
in the form of a Research Fellowship. EV was supported in part by NASA grant NNX08AH15G. We acknowledge the support 
of the KITP during the program 'Formation and Evolution of Globular Clusters', which was supported in part by the 
United States National Science Foundation under Grant No. PHY05-51164.

\label{lastpage}


\begin{thebibliography}{}

\bibitem[\protect\citeauthoryear{Allen, Moreno \& Pichardo}{2006}]{amp06}
Allen, C., Moreno, E., Pichardo, B., 2006, ApJ, 652, 1150

\bibitem[\protect\citeauthoryear{Balbinot et al.}{2009}]{baetal09}
Balbinot, E., Santiago, B.X., Bica, E., Bonatto, C., 2009, MNRAS, 396, 1596

\bibitem[\protect\citeauthoryear{Bastian \& Goodwin}{2006}]{bg06}
Bastian, N., Goodwin, S.P., 2006, MNRAS, 369, L9

\bibitem[\protect\citeauthoryear{Bastian et al.}{2008}]{betal08}
Bastian, N., et al., 2008, MNRAS, 389, 223

\bibitem[\protect\citeauthoryear{Baumgardt, Hut \& Heggie}{2002}]{bhh02}
Baumgardt, H., Hut, P., Heggie, D.C., 2002, MNRAS 336, 1069

\bibitem[\protect\citeauthoryear{Baumgardt \& Kroupa}{2007}]{bk07}
Baumgardt, H., Kroupa, P., 2007, MNRAS 380, 1589 

\bibitem[\protect\citeauthoryear{Baumgardt \& Makino}{2003}]{bm03}
Baumgardt, H., Makino, J., 2003, MNRAS 340, 227

\bibitem[\protect\citeauthoryear{Baumgardt, Makino \& Ebisuzaki}{2004}]{bme04}
Baumgardt, H., Makino, J., Ebisuzaki, T., 2004, ApJ 613, 1143

\bibitem[\protect\citeauthoryear{Baumgardt et al.}{2009}]{betal09}
Baumgardt, H., et al., 2009, MNRAS 396, 2051 

\bibitem[\protect\citeauthoryear{Bellazzini et al.}{2003}]{betal03}
Bellazzini, M., Ibata, R., Ferraro, F. R., Testa, V., 2003, A\&A 405, 577 

\bibitem[\protect\citeauthoryear{Bonatto \& Bica}{2008}]{bb08}
Bonatto, C., Bica, E., 2008, MNRAS, 479, 741 

\bibitem[\protect\citeauthoryear{Casetti-Dinescu et al.}{2007}]{detal07}
Casetti-Dinescu, D.I., Girard, T.M., Herrera, D., van Altena, W.F., L\'opez, C.E., Castillo, D.J., 2007, AJ, 134, 195

\bibitem[\protect\citeauthoryear{Chernoff \& Weinberg}{1990}]{cw90}
Chernoff, D.F., Weinberg, M.D., 1990, ApJ, 351, 121

\bibitem[\protect\citeauthoryear{Da Costa \& Armandroff}{1995}]{da95}
Da Costa, G.S., Armandroff, T.E., 1995, AJ, 109, 2533

\bibitem[\protect\citeauthoryear{Da Costa et al.}{2009}]{dc09}
Da Costa, G. S., Grebel, E. K., Jerjen, H., Rejkuba, M., Sharina, M. E., 2009, AJ in press, arXiv:0903.0215 

\bibitem[\protect\citeauthoryear{Dehnen et al.}{2004}]{detal04}
Dehnen, W., Odenkirchen, M., Grebel, E.K., Rix, H.W., 2004, AJ, 127, 2753

\bibitem[\protect\citeauthoryear{Dinescu et al.}{1999}]{detal99}
Dinescu, I.D., Girard, T.M., van Altena, W.F., 1999, AJ, 117, 1792

\bibitem[\protect\citeauthoryear{Elmegreen}{2008}]{e08}
Elmegreen, B.G., 2008, ApJ, 672, 1006

\bibitem[\protect\citeauthoryear{Fukushige \& Heggie}{1995}]{fh95}
Fukushige, T., Heggie, D.C., 1995, MNRAS, 276, 206

\bibitem[\protect\citeauthoryear{Gao et al.}{1991}]{getal91}
Gao, B., Goodman, J., Cohn, H., Murphy, B., 1991, ApJ, 370, 567

\bibitem[\protect\citeauthoryear{Gieles \& Baumgardt}{2008}]{gb08}
Gieles, M., Baumgardt, H., 2008, MNRAS, 389, 28

\bibitem[\protect\citeauthoryear{Giersz \& Heggie}{1994}]{gh94}
Giersz, M., Heggie, D.C., 1994, MNRAS, 270, 298

\bibitem[\protect\citeauthoryear{Goodman}{1984}]{g84}
Goodman, J., 1984, ApJ, 280, 298

\bibitem[\protect\citeauthoryear{G\"urkan et al.}{2004}]{gfr04}
G\"urkan, M.A., Freitag, M., Rasio, F.A., 2004, ApJ, 604, 632

\bibitem[\protect\citeauthoryear{Harris}{1996}]{h96}
Harris, W.E., 1996, AJ, 112, 487

\bibitem[\protect\citeauthoryear{Heggie et al.}{2006}]{hetal06}
Heggie, D.C., Trenti, M., Hut, P., 2006, MNRAS, 368, 677

\bibitem[\protect\citeauthoryear{Hills}{1980}]{h80}
Hills, J.G., 1980, ApJ, 235, 986

\bibitem[\protect\citeauthoryear{Ibata, Gilmore \& Irwin}{1994}]{igi94}
Ibata, R. A., Gilmore, G., Irwin, M. J., 1994, Nature, 370, 194

\bibitem[\protect\citeauthoryear{Innanen, Harris \& Webbink}{1983}]{ihw83}
Innanen, K.A., Harris, W.E., Webbink, R.F., 1983, AJ, 88, 338 

\bibitem[\protect\citeauthoryear{Jordi et al.}{2009}]{jetal09}
Jordi, K., Hilker, M., Baumgardt, H., Frank, M., Kroupa, P., Haghi, H., C\^ot\'e, P., Djorgovski, S. G., 2009, AJ, 137, 4586 

\bibitem[\protect\citeauthoryear{King}{1962}]{k62}
King, I., 1962, AJ, 67, 471

\bibitem[\protect\citeauthoryear{King}{1966}]{k66}
King, I., 1966, AJ, 71, 64

\bibitem[\protect\citeauthoryear{Kroupa}{2005}]{k05}
Kroupa, P., 2005, {\it The Fundamental Building Blocks of Galaxies}, in 
 Proceedings of the Gaia Symposium "The Three-Dimensional Universe with Gaia", Turon, C., O'Flaherty, K.S.,
  Perryman, M.A.C., eds., p.\ 629, astro-ph/0412069

\bibitem[\protect\citeauthoryear{K\"upper, Kroupa \& Baumgardt}{2008}]{kkb08}
K\"upper, A.H.W., Kroupa, P., Baumgardt, H., 2008, MNRAS, 389, 889

\bibitem[\protect\citeauthoryear{Lada \& Lada}{2003}]{ll03}
Lada, C. J., Lada, E. A., 2003, ARA\&A, 41, 57 

\bibitem[\protect\citeauthoryear{Larsen \& Brodie}{2000}]{lb00}
Larsen, S. S., Brodie, J. P., 2000, AJ, 120, 2938

\bibitem[\protect\citeauthoryear{Mackey \& Gilmore}{2004}]{mg04}
Mackey, A. D., Gilmore, G. F., 2004, MNRAS, 355, 504

\bibitem[\protect\citeauthoryear{Mackey \& van den Bergh}{2005}]{mv05}
Mackey, A. D., van den Bergh, S., 2005, MNRAS, 360, 631

\bibitem[\protect\citeauthoryear{Mackey et al.}{2007}]{metal07}
Mackey, A. D., Wilkinson, M. I., Davies, M. B., Gilmore, G. F., 2007, MNRAS, 379, 40 

\bibitem[\protect\citeauthoryear{Mackey et al.}{2008}]{metal08}
Mackey, A. D., Wilkinson, M. I., Davies, M. B., Gilmore, G. F., 2008, MNRAS, 386, 65

\bibitem[\protect\citeauthoryear{Martinez-Delgado et al.}{2002}]{metal02}
Martinez-Delgado, D., Zinn, R., Carrera, R., Gallart, C., 2002, ApJ, 573, 19

\bibitem[\protect\citeauthoryear{McLaughlin \& van der Marel}{2005}]{mm05}
McLaughlin, D. E., van der Marel, R. P., 2005, ApJ Suppl.\ Ser., 161, 304 

\bibitem[\protect\citeauthoryear{McMillan et al.}{1990}]{metal90}
McMillan, S., Hut, P., Makino, J., 1990, ApJ, 362, 522

\bibitem[\protect\citeauthoryear{Merritt et al.}{2004}]{metal04}
Merritt, D., Piatek, S., Portegies Zwart, S., Hemsendorf, M., 2004, ApJ 608, L25

\bibitem[\protect\citeauthoryear{Odenkirchen et al.}{2001}]{oetal01}
Odenkrichen, M., et al., 2001, ApJ, 548, L165

\bibitem[\protect\citeauthoryear{Parmentier et al.}{2000}]{petal00}
Parmentier, G., Jehin, E., Magain, P., Noels, A., Thoul, A.A., 2000, A \& A, 363, 526

\bibitem[\protect\citeauthoryear{Parmentier \& Grebel}{2005}]{pg05}
Parmentier, G., Grebel, E.K., 2005, MNRAS 359, 615

\bibitem[\protect\citeauthoryear{Parmentier}{2009}]{pg09}
Parmentier, G., 2009, in {\it Reviews in Modern Astronomy 21}, ed. S. R\"oser, Wiley-VCH, in press, arXiv:0901.3140

\bibitem[\protect\citeauthoryear{Peng et al.}{2006}]{petal06}
Peng, E., et al., 2006, ApJ, 639, 838

\bibitem[\protect\citeauthoryear{Pfalzner}{2009}]{p09}
Pfalzner, S., 2009, A\&A in press, arXiv:0904.0523v1 

\bibitem[\protect\citeauthoryear{Spitzer}{1987}]{s87}
Spitzer L.\ Jr., 1987, Dynamical Evolution of Globular Clusters,
  Princeton University Press, Princeton

\bibitem[\protect\citeauthoryear{Vesperini \& Heggie}{1997}]{vh97}
Vesperini, E., Heggie, D. C., 1997, MNRAS, 289, 898 

\bibitem[\protect\citeauthoryear{Vesperini \& Zepf}{2003}]{vz03}
Vesperini, E., Zepf, S. E., 2003, ApJ, 587, L97

\bibitem[\protect\citeauthoryear{Vesperini et al.}{2009}]{vetal09}
Vesperini, E., McMillan, S.L.W., Portegies Zwart, S., 2009, ApJ, 698, 615 

\bibitem[\protect\citeauthoryear{van den Bergh}{1993}]{vdb93}
van den Bergh, S., 1993, ApJ 411, 178 

\bibitem[\protect\citeauthoryear{Wilson}{1975}]{w75}
Wilson, C.P., 1975, AJ, 80, 175

\bibitem[\protect\citeauthoryear{Zinn}{1993}]{z93}
Zinn, R., 1993, in {\it The globular clusters-galaxy connection}, ASP Conf. Ser. 48, G. H. Smith and
 J. P. Brodie eds., p.\ 38

\end{thebibliography}
\end{document}